\PassOptionsToPackage{unicode}{hyperref}
\PassOptionsToPackage{hyphens}{url}
\PassOptionsToPackage{dvipsnames,svgnames,x11names}{xcolor}
\documentclass[
  12pt]{article}

\usepackage{amsmath,amssymb}
\usepackage{iftex}
\ifPDFTeX
  \usepackage[T1]{fontenc}
  \usepackage[utf8]{inputenc}
  \usepackage{textcomp} 
\else 
  \usepackage{unicode-math}
  \defaultfontfeatures{Scale=MatchLowercase}
  \defaultfontfeatures[\rmfamily]{Ligatures=TeX,Scale=1}
\fi
\usepackage{lmodern}
\ifPDFTeX\else  
\fi
\IfFileExists{upquote.sty}{\usepackage{upquote}}{}
\IfFileExists{microtype.sty}{
  \usepackage[]{microtype}
  \UseMicrotypeSet[protrusion]{basicmath} 
}{}
\makeatletter
\@ifundefined{KOMAClassName}{
  \IfFileExists{parskip.sty}{%
    \usepackage{parskip}
  }{
    \setlength{\parindent}{0pt}
    \setlength{\parskip}{6pt plus 2pt minus 1pt}}
}{
  \KOMAoptions{parskip=half}}
\makeatother
\usepackage{xcolor}
\makeatletter
\ifx\paragraph\undefined\else
  \let\oldparagraph\paragraph
  \renewcommand{\paragraph}{
    \@ifstar
      \xxxParagraphStar
      \xxxParagraphNoStar
  }
  \newcommand{\xxxParagraphStar}[1]{\oldparagraph*{#1}\mbox{}}
  \newcommand{\xxxParagraphNoStar}[1]{\oldparagraph{#1}\mbox{}}
\fi
\ifx\subparagraph\undefined\else
  \let\oldsubparagraph\subparagraph
  \renewcommand{\subparagraph}{
    \@ifstar
      \xxxSubParagraphStar
      \xxxSubParagraphNoStar
  }
  \newcommand{\xxxSubParagraphStar}[1]{\oldsubparagraph*{#1}\mbox{}}
  \newcommand{\xxxSubParagraphNoStar}[1]{\oldsubparagraph{#1}\mbox{}}
\fi
\makeatother

\usepackage{longtable,booktabs,array}
\usepackage{calc} 
\usepackage{etoolbox}
\makeatletter
\patchcmd\longtable{\par}{\if@noskipsec\mbox{}\fi\par}{}{}
\makeatother
\IfFileExists{footnotehyper.sty}{\usepackage{footnotehyper}}{\usepackage{footnote}}
\makesavenoteenv{longtable}
\usepackage{graphicx}
\makeatletter
\def\maxwidth{\ifdim\Gin@nat@width>\linewidth\linewidth\else\Gin@nat@width\fi}
\def\maxheight{\ifdim\Gin@nat@height>\textheight\textheight\else\Gin@nat@height\fi}
\makeatother
\setkeys{Gin}{width=\maxwidth,height=\maxheight,keepaspectratio}
\makeatletter
\def\fps@figure{htbp}
\makeatother

\makeatletter
\@ifpackageloaded{caption}{}{\usepackage{caption}}
\AtBeginDocument{%
\ifdefined\contentsname
  \renewcommand*\contentsname{Table of contents}
\else
  \newcommand\contentsname{Table of contents}
\fi
\ifdefined\listfigurename
  \renewcommand*\listfigurename{List of Figures}
\else
  \newcommand\listfigurename{List of Figures}
\fi
\ifdefined\listtablename
  \renewcommand*\listtablename{List of Tables}
\else
  \newcommand\listtablename{List of Tables}
\fi
\ifdefined\figurename
  \renewcommand*\figurename{Figure}
\else
  \newcommand\figurename{Figure}
\fi
\ifdefined\tablename
  \renewcommand*\tablename{Table}
\else
  \newcommand\tablename{Table}
\fi
}
\@ifpackageloaded{float}{}{\usepackage{float}}
\floatstyle{ruled}
\@ifundefined{c@chapter}{\newfloat{codelisting}{h}{lop}}{\newfloat{codelisting}{h}{lop}[chapter]}
\floatname{codelisting}{Listing}

\makeatother
\makeatletter
\@ifpackageloaded{caption}{}{\usepackage{caption}}
\@ifpackageloaded{subcaption}{}{\usepackage{subcaption}}
\makeatother

\ifLuaTeX
  \usepackage{selnolig}  
\fi
\usepackage{natbib}
\usepackage{har2nat} 
\defcitealias{hampson2022improving}{Hampson et al. 2022}

\usepackage{bookmark}

\IfFileExists{xurl.sty}{\usepackage{xurl}}{} 
\hypersetup{
  pdftitle={Title},
  pdfauthor={Author 1; Author 2},
  pdfkeywords={3 to 6 keywords, that do not appear in the title},
  colorlinks=true,
  linkcolor={blue},
  filecolor={Maroon},
  citecolor={Blue},
  urlcolor={Blue},
  pdfcreator={LaTeX via pandoc}}

\newcommand{\anon}{1}
\usepackage{pdflscape}
\usepackage{afterpage} 
\usepackage{tabularx}
\usepackage{ragged2e}
\usepackage{array}
\usepackage{booktabs}
\usepackage{authblk}

\begin{document}

\def\spacingset#1{\renewcommand{\baselinestretch}%
{#1}\small\normalsize} \spacingset{1}

\title{\bf Statistical Methodology Groups in the Pharmaceutical Industry \ifnum\anon=1\thanks{On behalf of the EFSPI Statistical Methodology Leaders Group}\fi}
\ifnum\anon=1
  \author[1]{Jenny Devenport \thanks{Corresponding author: jenny.devenport@roche.com}}
  \author[2]{Tobias Mielke}
  \author[3]{Mouna Akacha}
  \author[4]{Kaspar Rufibach}
  \author[1]{Alex Ocampo} 
  \author[5]{Vivian Lanius}
  \author[6]{Marc Vandemeulebroecke}
  \author[7]{Philip Hougaard}
  \author[8]{Pierre Collin}
  \author[9]{David Wright}
  \author[10]{Jurgen Hummel}
  \author[11]{Cornelia Ursula Kunz}
  \author[12]{Mike Krams} 

  \affil[1]{Hoffmann-La Roche AG, Switzerland}
  \affil[2]{Johnson and Johnson, Germany}
  \affil[3]{Novartis AG, Switzerland}
  \affil[4]{Merck KGaA, Switzerland}
  \affil[5]{UCB Biosciences GmbH, Germany}
  \affil[6]{Bayer BCC, Switzerland}
  \affil[7]{Lundbeck, Denmark}
  \affil[8]{Bristol Myers Squibb, Switzerland}
  \affil[9]{AstraZeneca UK Limited, UK}
  \affil[10]{Cytel, UK}
  \affil[11]{Boehringer Ingelheim International GmbH, Germany}
  \affil[12]{Berry Consultants, Austria}
\else
  \author{Authors Withheld for Blind Review}
\fi

\maketitle

\begin{abstract}
Research and Development is the largest budget position in the pharmaceutical industry, with clinical trials being a critical, yet costly and time-consuming component to inform decisions.  Beyond drug efficacy, the probability of success and efficiency of research and development are highly dependent on the approaches used for designing, analyzing, and interpreting clinical trials. Deep understanding of statistical methodology and quantitative approaches is therefore essential. Consequently, dedicated methodology groups have emerged in mid-size and large pharmaceutical companies and CROs. Their remit is to lead the conception and implementation of innovative quantitative methodologies in order to improve drug development, often by addressing complexities or offering more efficient designs. To achieve this, they collaborate internally and externally (e.g., with academics, regulators) to identify common challenges and tear down silos in order to invest in methods with the highest impact on efficiency and value to the portfolio.  Given the immense financial stakes of drug development—where delays carry massive implications—these groups represent a critical strategic investment. However, to realize this business impact, statistical innovations must be rigorously validated and seamlessly integrated.  This manuscript explores the setup, remit, and value of dedicated methodology groups, alongside the critical organizational considerations and success factors required to maximize their impact on the speed, efficiency, and probability of success.
\end{abstract}

\noindent%
{\it Keywords: Innovation, Design, Decision Making} 
\vfill

\newpage
\spacingset{1.8} 

\section{Introduction}\label{sec-intro}
The need for innovation in the pharmaceutical industry has been well documented.  Success rates for new drug approvals have remained stagnant over time while attrition rates, development timelines, and costs per new drug have increased \citep{dimasi2016innovation,laermann2021innovation,dowden2019trends,schuhmacher2025benchmarking}. Despite massive advancements in target identification technology and increases in research and development investments, corresponding increases in the success rates (e.g., marketing authorizations) have remained elusive giving rise to terms like the "productivity paradox," "innovation crisis," and "Eroom's law" \citep{Gassmann2004,laermann2021innovation,scannell2012diag}.   In other words, drug development is becoming slower and more expensive over time.  Marginal efforts to manage this as merely an efficiency problem (e.g., attempting to place tighter controls on project costs and speed, reorganizations to ensure lean resourcing, shifting outsourcing to lower-cost regions, and introducing performance scorecards) have been unsuccessful in improving the ratio of spend per approval \citep{scannell2012diag}.  Thus the pharmaceutical industry has turned to new ways to advance the drug development paradigm.

As the costs of drug development are driven by high failure rates, the pharmaceutical industry is making substantial strategic detours to overcome them.  Changes in portfolio composition have been observed such as increased focus on "me-too" fast followers like the GLP-1s \citep{Zilstorff2025glp}, avoiding low margin products (e.g., antibiotics), and pursuing high niche markets like orphan diseases and targeted therapies in oncology \citep{KINCH2023103622, kinch2024wanted}.  These strategies allow companies to increase their chances of success by pursuing targets with proven mechanisms of action or by focusing on targets or diseases with less competition.  Large pharmaceutical companies have also made changes to reduce risks in early development where failure rates are highest, particularly with increased mergers, acquisitions, and licensing deals \citep{schuhmacher2023analysis}.  However, in the case of mergers and acquisitions, some have pointed out that these are almost always followed by layoffs which may result in a loss of the knowledge and efficiencies that made such organizations successful \citep{kinch2024wanted}.  In addition, licensing deals are getting increasingly competitive and "picked over," meaning that deals must occur at earlier stages of development and as a result, the risk reduction is no longer realized \citep{kinch2024wanted}.

Pharmaceutical companies are also turning to disruptive technologies to attempt efficiency gains on multiple fronts. In the area of drug discovery, companies have been using machine learning in quantitative structure-activity relationship (QSAR) modeling to predict biological activity, toxicity, and other properties for decades (tools familiar to statisticians like random forests, and support vector machines).  According to a recent review, tools like graphical neural nets and transformer architectures are being employed to complete advanced drug design and prediction tasks with more complex data structures and models \citep{ferreira2025ai}.  Modeling and simulation techniques such as in-silico and organ-on-a-chip are being used as suggested by regulators to reduce or replace animal testing and other preclinical work \citep{fda2025roadmap}.  Artificial Intelligence (AI) and predictive modeling algorithms are also being explored along with multi-modal data to help with clinical development program optimization.  Examples include approaches like using real world data (RWD) to help evaluate trial eligibility criteria and broaden the patient pool while minimizing the impact on the treatment effect (e.g., TrialPathfinder, \citep{liu2021evaluating}) and mining external data sources to generate external control arms to reduce enrollment requirements.  However, so far this has been more widely accepted in early phases of research to inform decision making rather than broadly across confirmatory trials to inform assessment of treatment benefit due to the inadequacy of the data for this purpose and/or the need for untestable assumptions \citep{russek2025supplementing}. 

Arguably, decision-making on potential drug candidates is \textit{the} key competence of any pharmaceutical company. Ideally, such decision-making integrates any relevant source of evidence --  for example clinical data (internally or externally generated), RWD, benchmarks and AI powered approaches -- into a proper quantitative framework using assurance \citep{o2005assurance} and related metrics \citep{spiegelhalter1986monitoring, kunzmann2021review}, to compute a probability of success \citep{hampson2022improving,  hampson2022new}. The pivotal role statisticians and quantitative scientists play here in modeling, interpreting and communicating opportunities and risks appears evident. Statisticians have further contributed specific methodology to help address the needs for efficiency, speed, and complexity in new medical technology.  For example, Bayesian approaches have been developed and are applied frequently in pediatrics \citep{travis2023perspectives} and rare disease \citep{garczarek2023bayesian} to reduce required sample sizes and even shorten timelines needed to accumulate sufficient evidence for internal decision making and (in some contexts) regulatory approval of new medicines \citep{Best03042025, fdabayesian2026}.  A large arsenal of adaptive designs help remove white space, update design elements, or stop trials early for overwhelming efficacy or futility based on accumulating data \citep{he2014practical, wassmer2016group, wassmer2025group}.  Master protocols, basket, and umbrella designs facilitate the testing of multiple drugs in one or more patient populations simultaneously \citep{woodcock2017master,stallard2020}.  Statisticians are also contributing to the development of novel, composite endpoints to address ceiling effects or complex response patterns encountered by advanced therapies \citep{kappos2025composite}. Statisticians remain pivotal in making trial objectives more precise through the estimand framework \citep{ICHG2019E9r1,lipkovich2020causal}.  Beyond their quantitative expertise, statisticians use methodical, data-driven thinking to help shape overarching drug development strategies.

In fact, the demand for quantitative innovation continues to grow in the pharmaceutical industry as there is no shortage of problems for which critical thinking and statistics may offer solutions.  But to add value, these solutions must be implemented at scale \citep{rufibach2025implementation}.  In this paper we will discuss how dedicated methodology groups within pharmaceutical companies innovate and add value to their own organizations and beyond to inform scientific and regulatory discourse.  Section 2 clarifies the purpose and remit of methodology groups with examples from several pharmaceutical companies.  Section 3 reviews critical organizational considerations for dedicated methodology groups.  Sections 4 and 5 review internal and external factors that promote success and accelerate impact, respectively.  Section 6 characterizes the profiles of methodologists and characteristics that drive their success.  Finally, Section 7 provides a discussion and the future outlook for methodology groups.

\section{The remit of methodology groups}\label{sec-remit}

While differences in organizational structure, size, and specific tactics exist, in principle methodology groups in the pharmaceutical industry lead the conception, evaluation, implementation, and application of appropriate statistical methods.  This involves problem identification, devising and/or investigating potential solutions, and enabling appropriate adoption at scale through deployment of tools, templates, and ongoing education.  Additionally, the remit of methodologists naturally extends beyond theoretical development to include hands-on project consultation--supporting teams with study design, complex modeling, and regulatory responses.   To achieve these objectives and generate broad impact requires: 
\begin{itemize}
\item integration and collaboration with drug development teams
\item dedicated time for methodological work
\item connection and outreach to the greater statistical community
\item commitment to facilitate optimal implementation and broad adoption of methods.
\end{itemize}

We now discuss each of these essential activities in turn.  Methodology development in the pharmaceutical industry is rooted in identifying and addressing opportunities and challenges with impact to the core business of drug development. As such, integration and collaboration with drug development teams is crucial for methodology groups in order to identify and focus on relevant problems and understand practical requirements. Put in another way, methods groups cannot afford to reside in their "Ivory Tower" lest they fail to solve problems faced by development teams or offer solutions that are too complex to implement. Instead, they need to be familiar with the portfolio, the programs that are active in development, and the overarching strategy and the hurdles that are present in those programs. Successful methodological leaders need to summon their social skills and interact with the business directly, to offer guidance and increase awareness on the available internal expertise. Meetings with project team members (including but not only statisticians) are a must, either formally through consultations and informally in order to build up the trust and enable an open discussion of their programs. Offering trainings, sponsoring knowledge sharing events, and organizing hackathons \citep{bornkamp2024predicting} are also very effective ways to increase connectivity with drug development teams and share knowledge while promoting methodological excellence and opening opportunities for follow-up interaction. Implemented at scale, such involvement offers the additional advantage for an organized roll-out of statistical strategy across the organization.  Ideally, methodologists also contribute to functional governance/advisory boards, not to serve as "policeman," but to ensure that they can weigh in on key design and analysis features early –before decisions are final.  Lastly, independent expert input in governance/advisory boards introduces orthogonal thinking from those experts involved from within the projects and therapeutic leadership, thereby increasing either robustness of existing plans, or introducing novel approaches to advance outcomes via increases in speed, efficiency, or probability of success.  

This leads to the second requirement, dedicated time.  Developing and scaling novel methods to solve complex business problems cannot simply be squeezed in between accelerated project milestones. When methods development is treated as a secondary task, solutions evolve too slowly or fail to be rigorously tested for broader applications outside of an immediate project \citep{heinze2024phases}. To illustrate why dedicated time is useful, consider the following nuanced problems that directly impact trial conduct and success, which is where dedicated methodologists provide critical value:
\begin{itemize}
    \item Solving Complex Data Challenges: When facing the common business problem of missing data due to trial dropouts, methodologists had the dedicated time to partner with project teams to develop "reference-based multiple imputation" (rbmi) for application in clinical trials with longitudinal continuous endpoints. Rather than applying a rushed, one-off fix, they thoroughly evaluated the method, built specialized software, and published manuscripts and vignettes. This transformed an isolated project solution into a robust tool \citep{gower2022rbmi,wolbers2022standard}.
    \item Rescuing Trial Power in Immuno-Oncology (e.g., Non-Proportional Hazards): Traditional cancer trials rely on standard statistical models (such as the log-rank test) that are most powerful when a drug’s benefit remains constant over time. However, modern immunotherapies often take months to activate the immune system, leading to a delayed clinical effect. Dedicated methodologists have spearheaded the adoption of advanced survival analysis techniques to capture the clinical benefit of immunotherapies \citep{roychoudhury2023robust}.
\end{itemize}

Thirdly, connection to the greater statistics community including academia, statisticians at other pharmaceutical companies, and statisticians and other personnel at regulatory agencies enhances the efficiency, quality, and rigor of methodological innovation by aligning research or method dissemination agendas. This can take on a spectrum of activities including co-teaching classes, sponsoring / supervising PhD students, co-authoring papers, and co-sponsoring events (such as seminars, conferences, webinars, and workshops). Think of the difference between reading an article and engaging in a debate with the authors--the former is somewhat passive while the latter is quite active. From a knowledge-building perspective, active engagement facilitates progress more effectively, quickly unearthing gaps, biases, and opportunities for progress. This is precisely the motivation behind gatherings like the Statisticians in the Pharmaceutical Industry (PSI) Conference, Basel Biometrics Society (BBS) seminars \citeyear{BBS_Past_Events}, the annual European Federation of Statisticians in the Pharmaceutical Industry (EFSPI) Regulatory Statistics Workshop \citeyear{EFSPI_RSW_Event} and its American counterpart - the American Statistical Association (ASA) Biopharmaceutical (BIOP) Section Regulatory-Industry Statistics Workshop \citeyear{ASA_RSW_Event}, meetings with the EMA Methodology Working Party (e.g., the external controls workshop in advance of the EMA guidance document preparation \citep{EMA_ExtControl_Event}), and the ISCB 2025 Meeting - Statistical Thinking Across Borders:  Building Bridges and Expanding Horizons in Clinical Biostatistics \citep{ISCB25_Event}. Similarly, the formation and active engagement of special interest groups (SIGs), such as those sponsored by ASA BIOP and PSI / EFSPI \citep{ASAbiop_2025_wg,PSI_esig}, play important roles in fostering connections across disciplines, organizations, and geographies.  Covering foundational topics like decision making, safety, and regulatory, to emerging and advanced topics like artificial intelligence and machine learning, estimands, causal inference, software engineering and many others; these groups can also play important roles in raising awareness of and advancing impactful methodologies.  This is what makes SIGs attractive to methodologists to participate in and/or lead.  

Finally, taking steps to ensure that new methods can be broadly adopted (aka scaling up or commercialization) is an important part of innovation, as only those inventions that are broadly used add value to the organization or the field \citep{rufibach2025implementation}. New methods should be exposed to peer review. This means socialization at conferences and publication in peer reviewed / non-predatory journals to ensure availability to others. Methods development also benefits from collaboration and publication among statisticians from multiple organizations (especially including regulators and academics), as this lends greater credibility to the solution, which is important when promoting novelty in a highly regulated environment like drug development.  Furthermore, it is now quite common that such publications will include or refer to open source code or software for others to use.  Such inclusion removes an important implementation barrier. Code sharing, a common practice for decades, is straightforward operationally. But open-source software packages further enable community support and joint development / validation of new methods \citep{sanchez2021best, Weber2021}. While new methodological developments are often motivated or initiated by experiences on particular drug development projects, publication and scaling of the method may not be prioritized due to competing tasks.  When methodologists help project statisticians with a method for a specific project, the burden of evaluation, generalization, and publication (as well as the credit) is shared; and an enormous and enduring alliance is formed. Furthermore scaling efforts should not be solely focused on fellow statisticians - other stakeholders involved in decision making need to be consulted in an ongoing fashion to make sure that their educational needs, pragmatic concerns, and barriers to use are addressed.  A prime example of this is the multi-institutional effort involved in popularizing the \textit{causal roadmap} for generating real world evidence - addressing both statistical and nonstatistical audiences \citep{dang2023causal, ho2023current}.  In essence then, scaling up methods adoption should be seen as creating movement toward a new practice and not only a series of steps to complete. 

While these fundamental pillars--collaboration, dedicated time, community engagement, and adoption--are universal, their operationalization varies across the ecosystem.  The scope and tactical focus of a methodology group are influenced by the organization's size, resources, and strategic model.  Table 1 characterizes how a range of organizations (as archetypes) approach methods development, consultation, education, collaboration, and outreach.  It is notable that within each type of organization, there are variations in exactly how the groups are staffed and how support is implemented.  For example, some groups consist mainly of very senior staff who devote a percentage of time to long-term strategic innovation in addition to consulting.  In other groups, there is a mix of senior and junior methodologists and the work may be allocated differently to each.  Companies have also tried a decentralized approach where statisticians come together from across the organization on a voluntary basis to explore specific topics in addition to their day jobs \citep{kulmann2016bic}.  Regardless of the exact composition, if successful in these endeavors, methodology groups can make the statistics function, project teams, and the organization more innovative and efficient.  But there is another side to the story, which we explore in the next section. 

\begin{landscape}
\spacingset{1.0} 
\begin{table}[p]
    \centering
    \caption{Operationalization of methodology group remit by organizational subtype}
    \label{tab:industry_services}
    
    \renewcommand{\arraystretch}{1.5} 
    \footnotesize 
    
    \newcolumntype{L}{>{\RaggedRight\arraybackslash}X}
    
    \begin{tabularx}{\linewidth}{l L L L L L}
        \toprule
        & \textbf{Methods Development} & \textbf{Consultation} & \textbf{Education} & \textbf{Collaboration} & \textbf{Outreach} \\ 
        \midrule
        
        \textbf{Academia} & \textbf{Theoretical / Variable:} High volume of innovation, though practical application varies & \textbf{Multi-channel advisory:} with institutional peers, external granting agencies, and industry partners & \textbf{Core Mandate:} focused on standard curriculum, student supervision, and specialized seminars & \textbf{Critical Networking:} Variable engagement & \textbf{Dissemination Focus:} Publications, presenting at academic conferences and invited speaking engagements. \\ 
        \addlinespace 
        
        \textbf{Small Biotech} & \textbf{Reactive / crisis-driven:} Driven by the specific needs of projects & \textbf{Internal Triage:} In-house statisticians offer immediate advice / triage complex issues for allocation to external consultants when needed & \textbf{Team-specific:} Education is highly targeted, focusing mainly on immediate needs of project teams & \textbf{Competitive / selective:} Collaboration is variable, often limited by competitive pressures or resource constraints. & \textbf{Individual initiative:} driven by individuals rather than corporate strategy. \\ 
        \addlinespace
        
        \textbf{Mid-size Pharma} & \textbf{Pipeline-focused:} Priorities align with specific therapeutic needs or application of innovation & \textbf{Targeted support:} Support driven by specific issues on teams, strategic partner for development programs & \textbf{Knowledge sharing:} Offers training sessions and maintains specialist groups & \textbf{Network-based:} Maintains key external relationships and participates in expert teams to discuss designs before approval & \textbf{Professional engagement:} Leads research and education within specific specialist groups \\ 
        \addlinespace
        
        \textbf{Big Pharma} & \textbf{Broad / scalable:} Covers near and far term topics and tool development to scale innovation & \textbf{Strategic advisory:} provides ``state-of-the art'' consultations and holistic ``end-to-end'' advice to internal teams & \textbf{Formalized Training:} Develops tools, apps, classes, mentoring programs to ``upskill'' the wider organization & \textbf{Systemic + regulatory:} Partners with regulators, cross functional teams, and/or external partners for impact & \textbf{Advocacy \& influence:} Drives advocacy and acceptance for new approaches. Builds global networks to draw in talent and collaborators \\ 
        \addlinespace
        
        \textbf{CRO} & \textbf{Market-driven:} Differentiate services in the market by earning position as ``industry leader'' in specialized areas & \textbf{External service:} strategic consulting is the core product with specialized expertise to address client needs & \textbf{Client education:} Facilitates interpretation and implementation of complex methods specifically for client teams & \textbf{Partnership-focused:} Forges strategic partnerships to facilitate implementation and support enterprise-level initiatives & \textbf{Thought leadership:} Education and advocacy for innovative methods to enhance reputation, attract clients and guide clinical developers \\ 
        \bottomrule
    \end{tabularx}
\end{table}
\end{landscape}

\section{Critical organizational considerations for methodology groups} \label{sec-objection}
The EFSPI Statistical Methodology Leaders group consists of representatives from over 25 mid- and large pharmaceutical companies and CROs across Europe, demonstrating that there is a demand within companies for investment into and exchange on statistical methodology.   But as the pharmaceutical industry continues to experiment with right-sizing and investing in technologies and capabilities for the future \citep{mixtalent2026}, it is important to consider strategic friction points or risks of having dedicated methodologists / methodology groups.  Figure 1 depicts potential C-suite (i.e., Chief Executives or other top leaders) and functional considerations for such specialty groups.  These generally can be placed in two categories:  1)  financial and operational risks and 2) stagnation and opportunity costs.  These manifest differently in the C-suite and functional groups.

\begin{figure}[h]
\includegraphics[width=1.0 \textwidth]{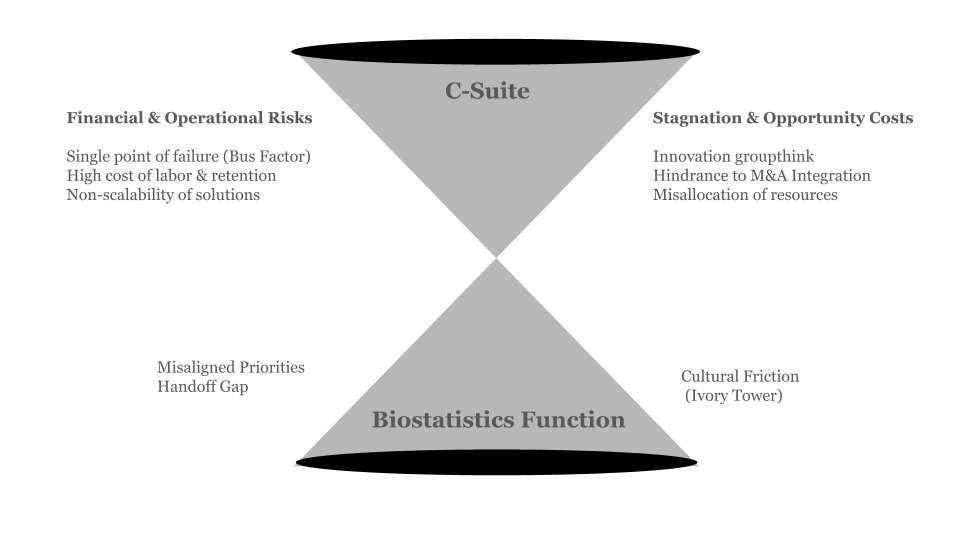}
\caption{C-suite and functional considerations for specialty groups}
\label{fig:figure1}
\end{figure}

The remit of the C-suite is to ensure the organization operates efficiently to achieve the timely approval and marketing of new medical products. However, specialty groups, whether in quantitative or medical functions, often present unique financial and operational challenges that can be difficult to quantify through standard metrics. The demographic composition of these groups sometimes trends toward highly skilled, senior-level experts. Unlike general project teams that may have a broader distribution of experience levels, specialty groups often function as concentrated centers of excellence. A high density of senior talent could result in higher collective costs compared to teams with more junior staff. Furthermore, the rarity of this expertise could create a significant "Bus Factor"\footnote{The Bus Factor refers to the risk to a project or business as a result of a team member or members suddenly not being available.}. If a specialist departs, they may take irreplaceable institutional knowledge and/or external networks with them, leading to concerns that such expertise is not easily scalable or transferable across the wider organization. While external consultants or academic partnerships may seem like a more flexible alternative, they introduce their other strategic considerations which will be addressed further below.  

A second category of concern involves potential stagnation or lost opportunity costs. This hypothetically manifests as "innovation groupthink," where a specialized group may inadvertently attract individuals with similar technical biases. If a methodology group collectively rejects a novel approach, valuable opportunities may be missed. Finally, in budget-constrained or lean environments, the value of a specialty group can be difficult to justify. Activities like highly specialized research or methodology development are not always readily visible in time-reporting systems as impacting a specific molecule. Especially during mergers and acquisitions, where leadership must rapidly identify essential talent, these groups risk being viewed as overhead rather than strategic drivers of long-term success.

Moving to the functional considerations about specialty groups, there are two similar but nuanced sets of concerns.  The first is around the misalignment of priorities.  When not actively consulting with a team on "their immediate challenge," the activities of the methodologist may not be seen as being inherently connected to the core business and consequently undervalued.   Lack of visibility of the method group's mission and priorities, a perceived disconnect between the group's day-to-day work and the work of study statisticians, and inadequate formal and informal interaction between the methodologists and study statisticians could fuel these ideas.  As alluded to with the "Ivory Tower" reference, an insufficient connection to the business or the perception of being "too forward looking" with respect to methodological problems may create the impression that the specialist group is tone-deaf to "real needs."  Another concern in this vein is the "hand-off gap."  When a potentially useful method or approach to problem solving is developed in the specialty group, there can be concerns that the resources needed to implement (tools or software, education, templates, etc.) are insufficient to enable the transfer of ownership back to the function.  In contrast to the Bus Factor and handover gap, there is a concern that separate specialty groups could make study statisticians feel less valued and cause retention problems within that resource pool. These concerns compound the story that specialty groups may create cultural friction that hinders the morale and outputs of the entire biostatistics function.    

While these considerations regarding specialized groups are notable, many organizations are operating successful methodology groups that actively and effectively mitigate these risks.  Methodology groups are intrinsically linked to the portfolio (via project consultancy) as well as to research and development strategy and efficiency (via participation in governance and collaborating to establish best practices). When well connected, the enterprise benefits are substantial.  Successes include portfolio-level futility analysis implementation, Bayesian decision-making frameworks \cite{dallow2018,hampson2022improving}, accelerated early oncology dose-finding strategies \cite{neuenschwander08}, model-based experimental designs \cite{bornkamp2009mcpmod}, and the widespread uptake of covariate adjustment to gain statistical efficiency and precision.  Methodology groups are expected to deliver scalable, high-value outcomes. By dedicating their focus to methodological advancement, these groups produce robust tools and frameworks that benefit multiple cross-functional teams simultaneously. Crucially, they bridge the internal "handover gap" by directly supporting project statisticians during implementation. They also drive external and internal scientific leadership through conference presentations, peer-reviewed publications, specialized training, and the sharing of reproducible code and software.  

The perception of "innovation groupthink" is readily addressed through strategic organizational design and hiring practices (see Section~\ref{sec:success}). Effectively functioning methodology groups are rarely composed entirely of senior personnel.  Integrating junior staff introduces recent technical training, fluency in modern programming languages, and fresh perspectives.  Because groupthink is fundamentally antithetical to innovation, cultivating a diverse team ensures the broader mindset necessary to tackle complex quantitative challenges. This diversity is amplified by fostering "subject matter expert" (SME) networks within and across companies, such as the estimand implementation working group \citep{EIWG_esig}, to facilitate training and promote co-ownership and the seamless adoption of novel methods.  Finally, to comment on the potential concern that methodology groups could create cultural friction: we firmly believe that the opposite is the case. A well-embedded and highly collaborative methodology group can elevate the impact and the morale of the broader quantitative enterprise.  

For small biotech companies or seeking to address only sporadic needs, outsourcing methodological and even functional work to external consultants and CROs is a common alternative to building dedicated teams.  The flexibility of outsourcing is a key benefit:  specialists are accessible on specific topics as needs arise.  For mid-sized and large pharmaceutical companies this strategy can also be useful to supplement internal needs in specific areas.  However, relying exclusively on this strategy perpetuates inherent structural risks.  First, a company still benefits from having sufficient internal expertise to evaluate whether a consultant’s proposal is truly fit-for-purpose and implementable within the internal operational systems. Second, outsourcing fundamentally fails to address the scalability problem. Consultants are typically retained for a fixed scope or duration; once the contract ends, subsequent advice or iterative troubleshooting necessitates a new contract. Consequently, the organization misses the opportunity to build critical in-house experience, replacing internal capability with external dependency.

Because quantifying methodological impact in standard "hard currency" metrics remains challenging, methodologists should continuously track, promote, and optimize their value proposition. Ultimately, the long-term success of these groups depends on their ability to clearly demonstrate how their core merits (informing efficient decisions, scaling knowledge, and bridging the handover gap) deliver compounding returns to the organization.

\section{Intrinsic factors that promote success} \label{sec-insuccess}
Companies that have built successful methodology groups contributing value internally and in the broader community share at least some intrinsic organizational characteristics.  First, methodology groups are integrated within the statistics organization. They are peers to the statisticians and naturally open for discussion and helping on applied problems. The integration is of key importance to ensure business impact, by understanding relevant problems and opportunities, and also by integrating project statisticians into relevant methodological initiatives.  Secondly, the groups are composed of diverse experts from various scientific / professional backgrounds (not all are necessarily statisticians), levels of seniority, and personality types. This diversity allows members to critically challenge and optimize innovative solutions. The different levels of seniority reduce arrogance and promote a fruitful blend of expert guidance, young ambition, and innovative mindset to yield an avalanche of ideas. Lastly, different types of personalities are required in methodological groups to combine the exactness of analytical science with clear communication of proposals to non-statisticians \citep{rufibach2025implementation}.

Beyond methodology group composition, the work culture needs to support exploration.  This is critical because invention does not lend itself to "production line" efficiency at first. In the exploration phase, there is at least some uncertainty and so the organization needs to tolerate ambiguity and experimentation--not forever, but for the discovery phase.  This creates space for reactive investigations and proactive (future-focused) opportunity finding.   During the exploration stage, methodologists should display curiosity and continuous learning that is fueled by scientific exchange, internally with interested colleagues, as well as externally through conferences, collaboration, and consultations.  In due course, priorities need to be selected and expected outcomes need to be delineated.  Finally, appropriate solutions should be brought for internal implementation and scaled through peer reviewed publications, conference presentations, and tools to facilitate implementation (e.g., training, templates, software).  This cycle, which moves from uncertainty to order, is important for finding the right problems and then devising solutions which can ultimately be adopted broadly (Figure 2).

\begin{figure}[h]
    \centering
    \includegraphics[width=1.0\textwidth]{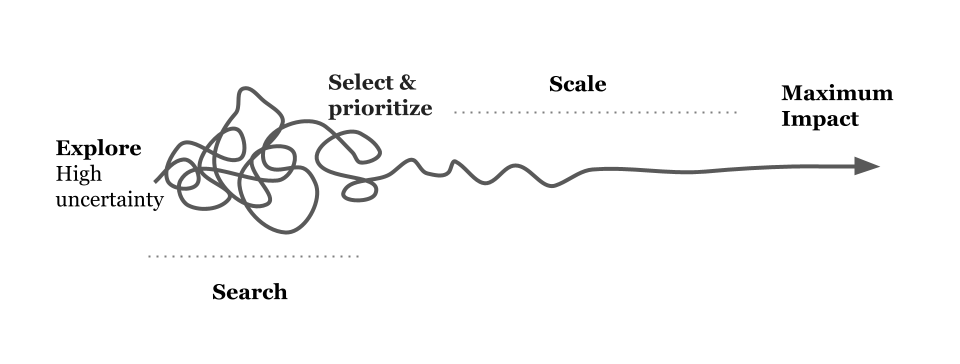}
    \caption{The innovation cycle in methodology groups}
    \label{fig:figure2}
\end{figure}

This cycle is also relevant in contexts where methodologists (either by volunteering or assignment) join and inform the content and outcomes of cross-functional initiatives.  Trying to interject statistical thinking in \textit{one-go} may not be as effective as partnering for the longer term, which allows for deeper understanding of the stakeholders, the obstacles, and how to best align statistical thinking with scientific and business objectives.  Examples of this are readily apparent as the pharmaceutical industry looks for opportunities to accelerate drug development through the use of RWD for creating external control arms and post marketing evidence generation and AI / ML for integrating information to support decisions.  While methodologists can instinctively articulate the risks of these approaches, their most vital contribution is to create and demonstrate processes for evaluation or validation of their use within particular contexts \citep{behr2025opportunities, fdarwd2023, fdaaireg2025}.  Ultimately, methodologists must actively champion the realization of these opportunities, balancing risk mitigation with drive for innovation, to be fully recognized as strategic partners.

Psychological safety, a construct strongly related to an exploration-tolerant organization, is generally recognized as an essential ingredient for innovation \citep{bonterre2025ps}.  The nature of methodology work involves uncertainty, and even failure, which results in new learnings and future opportunities.  Psychological safety allows members to explore and take risks without fear of embarrassment or retribution. Conceptual progress in methodological work relies on the freedom to fail.  Without psychological safety, statisticians are denied the vulnerability required to challenge the status quo, which is formidable among statisticians and non-statisticians alike.  As recently summarized by Berry \citeyear{berry2025statistical}, "The 'right way' becomes frozen in lore. This is so even when the 'right way' is not best." Ultimately, it is the psychological safety created by individuals, leadership and the broader organization that enables methodologists to have very open dialog (especially, but not only with each other), to give and receive constructive feedback, and to offer mutual support.  This helps expedite innovation, cope with risks, and generate better solutions. 

Transparency and ease of communication between methodology groups and internal stakeholders is also needed.  Sharing key priorities via the preferred route of communication (a web portal, chatroom, or newsletter) is appreciated - particularly when it is accompanied by an invitation for feedback or further dialog.  This again helps with the perception that the specialty group is a part of the whole, rather than simply a distinct part.  Organizing regular trainings, question and answer sessions, consulting hours, and even informal chats also help bring the methodology experts closer to their communities and build new avenues of internal collaboration.   As previously described by \citet{rufibach2025implementation} the creation of web portals can facilitate broad consideration and use of regulatory documents, tutorials, training materials, examples, code, frequently asked questions, and reviews of new / most relevant literature on statistical methodology topics.   This certainly helps break down barriers to implementation of new methods and builds up trust.

Collaborators within the wider organization (other statisticians, software developers, and also other departments like clinical, operations, regulatory, portfolio, and IT) are particularly useful to identifying the right problems to solve, helping to evaluate potential solutions (feasibility and acceptability) and to generate or maintain momentum.  As already discussed, when methodologists are isolated from the core business they risk solving nonessential problems with solutions that may not be feasible. Moreover, wide adoption of novel methods requires advocacy from many places within the organization, not only the biostatistics function. Bringing others along for the journey, learning about their concerns and addressing them, and giving them some share in the solution and its implementation can facilitate support.  Implementation of estimands in regulatory clinical trials has certainly commanded cross-functional participation; and publications from the Estimand Implementation Working Group \citep{EIWG_esig} exemplify the benefits of collaborating to ensure that considerations from regulators, medical, and HTA stakeholders are addressed \citep{Vuong_2026estimands,lanius2025realizing}.

Last but not least, methodologists also need obvious and unobstructed paths for engagement with / endorsement by senior leaders within their organizations (ideally, formally through seats on leadership teams).  This support is particularly vital from leadership outside of the immediate statistics function--such as Heads of Data Science, Clinical Development, and Research and Development.  At an operational level, this helps ensure that the methodology work is perceived as valuable to the enterprise.  But strategically, connection with senior leaders is critical for timely allocation of resources when needed (e.g., as might be the case for short term contracts); and also for ensuring endorsement, prioritization, and implementation of novel methods and best practices for routine use.  Senior management has an \textit{architectural influence} on innovation through the opportunities and constraints they place on the environment:  whether this is encouragement of experimentation and tolerance of failure; the willingness to fund small demonstrations like hackathons or external partnerships with academic or nonprofit organizations; or sponsorship of major initiatives such as changes to software or hardware infrastructure \citep{cortes2021strategic}.

\section{External factors that accelerate impact:  the case for outreach and external collaboration}

While connection to the internal business is important, one can quickly see the innovator's dilemma in the pharmaceutical and other industries \citep{christensen2015innovator}:  "Blindly following the maxim that good [methodologists] should keep close to their customers can sometimes be a fatal mistake."\footnote{The original quote names "managers" instead of "methodologists"}  In this analogy from the microchip world, Christensen illustrated how businesses can get so focused on their own problems and their own ways of working that they fail to make necessary and obvious changes or consider easier solutions.  It is an important reminder that methodologists must pursue external insights, e.g. by building global networks that can prevent narrow thinking and accelerate progress.  Forums like Project Significant, sponsored by the FDA's Oncology Center of Excellence, are expressly designed to help forge connections and facilitate discussion on the design/analysis challenges in cancer clinical trials \citep{fda2025sig}.  

Global outreach and external collaborations with statisticians from other companies, academia, and regulators can seem like a delicate thing in the highly competitive, heavily regulated pharmaceutical industry.  However, if one accepts the notion that the intellectual property is the medicine or medical product, then there is a path forward for collaboration which may also aid in the acceptance of novel methods.  One needs to set a common understanding of what information can be shared without a nondisclosure agreement and stay within those limits.  Sometimes, no data sharing is involved, and statisticians from various organizations including regulatory agencies can work together to clarify challenges, such as aligning on terminology for interim analyses \citep{asikanius2025clinical}.  Other times, data must be exchanged, and here technology can help.  Federated learning tools where data are analyzed locally and only aggregate summaries are shared across companies can facilitate more thorough evaluation or demonstration of new methods.   The SAVVY consortium is one example where, united via the observation that the traditional reporting of raw incidence rates of adverse events could be quite biased by varying follow-up time and competing events, statisticians worked together to identify and test simple yet effective methods for improved estimation of event rates \citep{rufibach2024survival}.  Both of these examples, which came together rather quickly, further highlight the value and efficiency of collaborations to address common challenges and potentially facilitate or accelerate regulatory acceptance of novel methods in a manner greater than the stand-alone effort of one company.  

Building up the external visibility of the methodology group and its members can be valuable in other ways, including facilitating recruitment of new talent in the company--not only in the methods group, but in the broader quantitative organization.  By maintaining relationships with academia, supervising students (often of the academics one partners with), and participating in conferences, methodologists build positive regard for companies, making them more attractive as potential employers.  This can be mutually beneficial as students have built up positive relationships with other employees and may fit in better at the start of their employment due to their familiarity with the people, systems, and culture.  Many pharmaceutical companies and even some large medical institutions have statistical internship programs geared toward next-generation talent \citep{stattrak2026}.  In addition, these relationships with academia also help expose the academic researchers to the challenges and opportunities of most relevance to drug development, thereby guiding them on relevant research topics for doctoral students, grants, and other mutually beneficial collaborations.

Methodologists may also serve as technical liaisons in scientific dialogues with regulators, helping to navigate the nuances of emerging methods or technologies and address areas of uncertainty. By grounding these discussions in objective statistical and scientific principles rather than individual corporate interests, they act as credible partners who can identify gaps, case studies and educational paths that foster mutual alignment. This is key, because until companies deeply understand the root of regulator uncertainty, they cannot address those concerns across the spectrum of their portfolios.  These collaborative exchanges are often facilitated in curated forums such as statistical conferences (e.g., EFSPI and ASA Regulatory Statistics Workshops \citep{EFSPI_RSW_Event, ASA_RSW_Event}), public workshops such as the ones by the European Medicines Agency to discuss potential benefits and challenges of particular statistical approaches and align on useful information to include in upcoming guidance documents \citep{EMA_2024_mwp, EMA_ExtControl_Event}, or the FDA's Complex Innovative Trial Design program \citeyear{fda2025cidmp}.  

\section{Profiles of successful methodologists}
\label{sec:success}
Technical skills obviously top the list of qualities that one might consider for a methodology group member.  The question of what qualifies an individual to be a methodology expert is similar to the question of what qualifies a statistician in general.  As stated in ICH E6 and reiterated in ICH E9 \citep{ICHG2025E6, ICHG2025E9}, trial statisticians should be qualified by education, training, and experience.  Expertise, of course, implies high levels of technical capabilities.  However, there is no one route, no single recipe or alchemy, that allows one to arrive at the title of expert methodologist. There are highly effective methodologists who spent the first half of their careers in academia or in other industries and only later joined a specialty group in the pharmaceutical industry. Similarly, there are methodologists who acquired much of their skill on projects (e.g., as trial statisticians) due to their own astute powers of observation, technical curiosity, and willingness to obsess over the details, implications, and extensions long after their projects were completed. An EFSPI Working Group \citeyear{efspiwg1999qs} reached a similar conclusion about the qualifications of statisticians in general - especially noting the variability in degree programs in different European countries and lack of alignment on the criteria to support universal accreditation.  In 2025, the ASA BIOP \citep{asabiop2025stat} also highlighted how the demand for statisticians in all areas of drug development, even with the explosion of AI, continues to grow.  Methodologists may add value by supporting the assessment of such innovations and, if viable, facilitating implementation at scale via training, templates and processes.  

In order to understand drug development problems and be able to translate them into scientific research questions, match often imprecise clinical questions to appropriate methods and designs, and understand regulatory thinking, statistical methodologists should have a solid background and keen interest in drug development.  Familiarity with the clinical development process and regulatory guidelines is important, but not only from an academic, theoretical perspective. Having actual hands-on experience with implementation helps ensure that proposed solutions to statistical challenges are pragmatic and feasible.  It is also relevant to understand (through lived experience) the challenges of influencing a cross-functional team. Again, if one only works on problems in isolation, without the pressure of operational and medical details and restricted timelines, it is more difficult to anticipate headwinds to implementation of complex statistical innovations - even if they are well suited to the problem at hand. It also enhances the methodologist's credibility with experienced drug developers cross functionally (e.g., not only project lead statisticians and their management chain, but also medical and commercial team leaders, etc.) because they need to trust that they can have a reasonable exchange and debate with the methodologist that extends beyond technical details and into strategy.

As the demand for novel quantitative methods to address clinical development questions increases, the need for the methodology group to have additional technical capabilities and promotional skills increases. On the technical side, there is a growing need to include software development as part of the methods development cycle in order to ensure that the broader enterprise is able to readily implement the innovation. This coincides nicely with the open-source statistical software revolution currently ongoing in the pharmaceutical industry \citep{bove2026sse}. To allow software innovation to keep pace with advancing statistical methodologies, several approaches are being applied: some companies are employing statistical software engineers within their methodology / innovation groups directly, and others are utilizing a variety of options including but not limited to partnering with the IT department, contracting with developers, sponsoring students, or joining consortia as needed. Certainly, an advantage of open source packages (regardless of origin) is the opportunity for the community to offer feedback and share the responsibility for lifecycle management / maintenance going forward. Communication is also critical to scaling up and promoting the use of innovative methods. If the inventor of a novel method is not excited about nor adept at this aspect of innovation, then it is necessary to have others who can promote its use beyond the initial publication through seminars, invited presentations, podcasts, blogs, and other similar activities. Engaging with the community through education, case-study sharing, and debate is critical to the uptake and incorporation of novel methods into the standard toolkit of drug development.

Beyond subject matter expertise, there are the personality traits and so-called \textit{soft-skills} that can really help the methodologist thrive. Curiosity and intrinsic motivation are critical traits, prompting active engagement in new topics. This often means learning a new method by doing / applying it, teaching it, and/or starting a SIG with other experts and collaborators. As alluded to previously, understanding the value of finite and long-term goals is important for efficiency. No topic should be polished forever to reach perfection, and not every challenge can be resolved within one day - otherwise it would have been resolved long ago.  Appreciating the difference between scenarios calling for quick wins versus long-term collaborations is essential for success. Similarly, appreciating the difference between incremental, yet time-consuming perfections and transformational novel approaches to problem solving is of key importance. To achieve this, one needs touch points internally and externally to anticipate trends and capitalize on opportunities. This speaks again to the importance of outreach and collaboration--which also require a certain amount of sociability and the inclination to act. Methodologists should be able to engage without necessarily waiting for an invitation to a forum or meeting. While invitations are expedient, when they do not come and the topic is important, methodologists should be prepared to invite themselves or, when necessary, create their own forum. The same can be said for knowledge sharing and training within organizations and externally. Collaborators can usually be found in both places, lowering the hurdle to start and permitting shared workload and accomplishments.

Importantly, not all organizations prioritize these characteristics in the same way, and no single methodologist possesses all of these qualities and interests at the same time. This is determined in part by the organizational reporting structure (where the methodology "sits"), the leadership philosophy espoused by the organization (including but not limited to inclusion of methodologists in the leadership team and governance structure), the portfolio of the company, and the interests of the group. There is value in building diverse teams that collectively carry these qualities - that is, a diversity of technical expertise, drug-development experience, personality, and curiosity to drive statistical innovation and outstanding performance.  

\section{Discussion and future outlook} 

This manuscript provides an overview of the remit and value of statistical methodology groups in the pharmaceutical industry.  Methodology groups are not merely a luxury, but rather a strategic imperative.  The pharmaceutical industry has faced a well-documented innovation crisis for decades, where the cost per approval continues to rise.  Cost cutting measures (sourcing strategies and reorganizations) intended to address inefficiencies have yielded only marginal gains.  As the pharmaceutical and regulatory ecosystems continue to evolve with complex data sources and disruptive technologies, the need for robust yet innovative statistical methods to inform decision making increases at the trial, molecule, program, and portfolio levels.  Without dedicated personnel working on, anticipating, and understanding novel developments, organizations risk being merely reactive (addressing challenges as they arrive) or wasting resources on pseudo-innovation that yields only incremental benefits.  

Beyond theoretical exploration, statistical methodology groups have demonstrated tangible value in addressing this productivity paradox. By integrating evidence into quantitative frameworks, these groups enable organizations to assess probability of success and make better investment decisions. They have been pivotal in addressing the complexity of medical innovation through the application of Bayesian approaches in rare diseases, which can reduce sample sizes and shorten timelines. Additionally, the implementation of adaptive designs, master protocols, and the estimand framework have better aligned trial designs and analyses with research objectives and allowed development programs to become more efficient. These contributions highlight that methodological expertise is the engine that converts potential into rigorous, approvable scientific evidence.

Despite this value, concerns about methodology groups may occasionally surface: the risk of creating "Ivory Towers" that do not solve relevant business problems, or perhaps provide solutions that cannot realize the targeted value in the organization due to problems with scaling. However, these objections fall short when considering the alternative: using consultants to address challenges as they arise. Due to the nature of consultant contracts which are limited in scope and time, this model fails to build institutional knowledge and fully generalizable solutions. Dedicated methodology groups bridge the "handover gap" because in addition to collaboratively solving problems as they arise;  they create tools, software, training, and document templates that facilitate re-use of solutions across the entire organization.   They are also instrumental - through external outreach and collaboration - in facilitating that the broad community (including industry peers, academia and regulators) are familiar with and accepting of novel solutions. Innovation adds the most value when it is implemented at scale. Therefore, methodology groups serve as the essential infrastructure for scaling expert knowledge, ensuring that methods become standard practice rather than the special case.   

Realizing this potential requires specific organizational and individual considerations. Ideally, organizations foster a culture of "psychological safety," granting methodologists the freedom to explore in addition to the expectation of delivering value.  On an individual and group level, exceptional technical capabilities are the starting point. But soft skills like intrinsic motivation, business savvy, curiosity, and sociability are key to identifying opportunities, prioritizing the right projects (near and far term), and tearing down silos to ensure success. The most valuable methodologists are those who actively collaborate within and outside the business to solve problems, create collaborations, and drive changes in practice.  

The benefits of outreach and external collaboration cannot be overstated. The "innovator's dilemma" \citep{christensen2015innovator} suggests that organizations focusing solely on internal problems miss opportunities and risk becoming obsolete. To continue to realize value from quantitative innovation, organizations must continue to invest in methodology groups and their ability to form impactful collaborations. As demonstrated by the authors of this manuscript--methodology leaders from across the pharmaceutical industry and specialized CROs who are members of the EFSPI Statistical Methodology Leaders group \citeyear{statmethwg_esig}--there is increasing interest and benefit from working together. We identify priority topics and co-create pathways to drive the field forward. Ultimately, methodology groups work in partnership with other quantitative scientists and stakeholders to ensure scientific integrity in a rapidly changing environment. Whether through advancing stakeholder perspectives on novel methods or validating AI-driven insights and designs, they ensure that the industry's pursuit of efficiency and speed does not compromise the foundation of evidence generation. The examples and arguments presented herein should serve as a roadmap for statisticians to advocate for these groups as an essential component of the organizational blueprint for successful drug development in the future.    
\label{bibtex}

\bibliography{bibliography.bib}

@article{rufibach2025implementation,
  title={Implementation of statistical innovation in a pharmaceutical company},
  author={Rufibach, Kaspar and Wolbers, Marcel and Devenport, Jenny and Yung, Godwin and Harbron, Chris and Bedding, Alun and Huang, Zhiyue and Lin, Ray and Pang, Herbert and Saban{\'e}s Bov{\'e}, Daniel and others},
  journal={Statistics in Biopharmaceutical Research},
  volume={17},
  number={1},
  pages={113--124},
  year={2025},
  publisher={Taylor \& Francis}
}

@article{laermann2021innovation,
  title={Innovation crisis in the pharmaceutical industry? A survey},
  author={Laermann-Nguyen, Ute and Backfisch, Martin},
  journal={SN Business \& Economics},
  volume={1},
  number={12},
  pages={164},
  year={2021},
  month={dec},
  publisher={Springer},
  doi={10.1007/s43546-021-00163-5}
}

@article{travis2023perspectives,
  title = {Perspectives on informative {B}ayesian methods in pediatrics},
  author = {Travis, James and Rothmann, Mark and Thomson, Andrew},
  journal = {Journal of Biopharmaceutical Statistics},
  volume = {33},
  number = {6},
  pages = {830--843},
  year = {2023},
  publisher = {Taylor \& Francis}
}

@article{lipkovich2020causal,
  title={Causal inference and estimands in clinical trials},
  author={Lipkovich, Ilya and Ratitch, Bohdana and Mallinckrodt, Craig H},
  journal={Statistics in Biopharmaceutical Research},
  volume={12},
  number={1},
  pages={54--67},
  year={2020},
  publisher={Taylor \& Francis}
}

@book{wassmer2016group,
  title={Group sequential and confirmatory adaptive designs in clinical trials},
  author={Wassmer, Gernot and Brannath, Werner},
  volume={301},
  year={2016},
  publisher={Springer}
}

@book{wassmer2025group,
  title={Group sequential and confirmatory adaptive designs in clinical trials, second edition},
  author={Wassmer, Gernot and Brannath, Werner},
  volume={301},
  year={2025},
  doi = {10.1007/978-3-031-89669-9},
  publisher={Springer Cham}
}

@article{kappos2025composite,
  title={Composite confirmed disability worsening/progression is a useful clinical endpoint for multiple sclerosis clinical trials},
  author={Kappos, Ludwig and Yiu, Sean and Dahlke, Frank and Coetzee, Timothy and Cutter, Gary R and Yuen, Steven and Bonati, Ulrike and Lublin, Fred D},
  journal={Neurology},
  volume={104},
  number={10},
  pages={e213558},
  year={2025},
  publisher={Lippincott Williams \& Wilkins Hagerstown, MD}
}

@article{Weber2021,
  title   = {Applying Meta-Analytic-Predictive Priors with the {R} {B}ayesian Evidence Synthesis Tools},
  author  = {Sebastian Weber and Yue Li and John W. Seaman and Tomoyuki Kakizume and Heinz Schmidli},
  journal = {Journal of Statistical Software},
  year    = {2021},
  volume  = {100},
  number  = {19},
  pages   = {1--32},
  doi     = {10.18637/jss.v100.i19}
}

@article{scannell2012diag,
  title={{Diagnosing the decline in pharmaceutical R\&D efficiency}},
  author={Scannell, Jack W and Blanckley, Alex and Boldon, Helen and Warrington, Brian},
  journal={Nature Reviews Drug Discovery},
  volume={11},
  number={3},
  pages={191--200},
  year={2012},
  publisher={Nature Publishing Group UK London}
}

@article{schuhmacher2025benchmarking,
  title={Benchmarking R\&D success rates of leading pharmaceutical companies: an empirical analysis of FDA approvals (2006--2022)},
  author={Schuhmacher, Alexander and Hinder, Markus and Brief, Elazar and Gassmann, Oliver and Hartl, Dominik},
  journal={Drug Discovery Today},
  pages={104291},
  year={2025},
  publisher={Elsevier}
}

@article{kinch2024wanted,
  title={{WANTED DEAD OR ALIVE: New Thinking to Incentivize Drug Development}},
  author={Kinch, Michael S},
  journal={Pharmaceutical Research},
  volume={41},
  number={2},
  pages={199--202},
  year={2024},
  publisher={Springer}
}

@article{KINCH2023103622,
title = {{2022 in review: FDA approvals of new medicines}},
journal = {Drug Discovery Today},
volume = {28},
number = {8},
pages = {103622},
year = {2023},
doi = {https://doi.org/10.1016/j.drudis.2023.103622},
author = {Michael S. Kinch and Zachary Kraft and Tyler Schwartz},
}

@article{woodcock2017master,
  title={Master protocols to study multiple therapies, multiple diseases, or both},
  author={Woodcock, Janet and LaVange, Lisa M},
  journal={New England Journal of Medicine},
  volume={377},
  number={1},
  pages={62--70},
  year={2017},
  publisher={Mass Medical Soc}
}

@article{dowden2019trends,
  title={Trends in clinical success rates and therapeutic focus},
  author={Dowden, Helen and Munro, Jamie},
  journal={Nat Rev Drug Discov},
  volume={18},
  number={7},
  pages={495--496},
  year={2019}
}

@Inbook{Gassmann2004,
  author="Gassmann, Oliver and Reepmeyer, Gerrit and von Zedtwitz, Maximilian",
  title="Innovation as a Key Success Factor in the Pharmaceutical Industry",
  bookTitle="Leading Pharmaceutical Innovation: Trends and Drivers for Growth in the Pharmaceutical Industry",
  year="2004",
  publisher="Springer Berlin Heidelberg",
  address="Berlin, Heidelberg",
  pages="1--22",
  doi="10.1007/978-3-540-24781-4_1"
}

@article{dimasi2016innovation,
  title={{Innovation in the pharmaceutical industry: New estimates of R\&D costs}},
  author={DiMasi, Joseph A and Grabowski, Henry G and Hansen, Ronald W},
  journal={Journal of Health economics},
  volume={47},
  pages={20--33},
  year={2016},
  publisher={Elsevier}
}

@article{schuhmacher2023analysis,
  title={Analysis of pharma R\&D productivity--a new perspective needed},
  author={Schuhmacher, Alexander and Hinder, Markus and und Stein, Alexander von Stegmann and Hartl, Dominik and Gassmann, Oliver},
  journal={Drug Discovery Today},
  volume={28},
  number={10},
  pages={103726},
  year={2023},
  publisher={Elsevier}
}

@misc{Zilstorff2025glp,
  author = {Zilstorff, Pandora},
  title = {{Two companies poised to capitalize on the rise of GLP-1 drugs: Why Pfizer and Roche look like top choices for investors in the war to tip the scales}},
  year = {2025},
  month = {12},
  howpublished = {\url{https://www.morningstar.com/sustainable-investing/2-companies-poised-capitalize-rise-glp-1-drugs?utm_source=copilot.com}},
  note = {Accessed: 2025-12-12} 
}

@misc{stattrak2026,
  author = {{American Statistical Association}},
  title = {2026 Internships},
  year = {2026},
  month = {1},
  howpublished = {\url{https://stattrak.amstat.org/2025/12/01/2026-internships/#:~:text=We are looking for multiple,and analysis of clinical trials}},
  note = {Accessed: 2026-01-27} 
}

@article{gower2022rbmi,
  title={{rbmi: An R package for standard and reference-based multiple imputation methods}},
  author={Gower-Page, Craig and Noci, Alessandro and Wolbers, Marcel},
  journal={Journal of Open Source Software},
  volume={7},
  number={74},
  pages={4251},
  year={2022}
}

@misc{mixtalent2026,
  title= {7 talent trends for the life sciences 2026},
  author={{Mix Talent}},
  year = {2026},
  howpublished = {\url{https://mix-talent.com/trends-report-2026/}},
  note = {Accessed: 2026-01-27} 
}

@article{wolbers2022standard,
  title={Standard and reference-based conditional mean imputation},
  author={Wolbers, Marcel and Noci, Alessandro and Delmar, Paul and Gower-Page, Craig and Yiu, Sean and Bartlett, Jonathan W},
  journal={Pharmaceutical statistics},
  volume={21},
  number={6},
  pages={1246--1257},
  year={2022},
  publisher={Wiley Online Library}
}

@article{sanchez2021best,
  title={Best practices in statistical computing},
  author={Sanchez, Ricardo and Griffin, Beth Ann and Pane, Joseph and McCaffrey, Daniel F},
  journal={Statistics in medicine},
  volume={40},
  number={27},
  pages={6057--6068},
  year={2021},
  publisher={Wiley Online Library}
}

@misc{bonterre2025ps,
  author={Bonterre, Michelle},
  title={Why psychological safety is the hidden engine behind innovation and transformation },
    institution ={Harvard Business Impact Insights} ,
  howpublished = {\url{https://www.harvardbusiness.org/insight/why-psychological-safety-is-the-hidden-engine-behind-innovation-and-transformation/}},
year = {2025},
  note = {Accessed: 2025-12-16} 
}

@article{o2005assurance,
  title={Assurance in clinical trial design},
  author={O'Hagan, Anthony and Stevens, John W and Campbell, Michael J},
  journal={Pharmaceutical Statistics: The Journal of Applied Statistics in the Pharmaceutical Industry},
  volume={4},
  number={3},
  pages={187--201},
  year={2005},
  publisher={Wiley Online Library}
}

@article{hampson2022improving,
  title={Improving the assessment of the probability of success in late stage drug development},
  author={Hampson, Lisa V and Bornkamp, Bjoern and Holzhauer, Bjoern and Kahn, Joseph and Lange, Markus R and Luo, Wen-Lin and Cioppa, Giovanni Della and Stott, Kelvin and Ballerstedt, Steffen},
  journal={Pharmaceutical Statistics},
  volume={21},
  number={2},
  pages={439--459},
  year={2022a},
  publisher={Wiley Online Library}
}

@article{hampson2022new,
  title={A new comprehensive approach to assess the probability of success of development programs before pivotal trials},
  author={Hampson, Lisa V and Holzhauer, Bj{\"o}rn and Bornkamp, Bj{\"o}rn and Kahn, Joseph and Lange, Markus R and Luo, Wen-Lin and Singh, Pritibha and Ballerstedt, Steffen and Cioppa, Giovanni Della},
  journal={Clinical Pharmacology \& Therapeutics},
  volume={111},
  number={5},
  pages={1050--1060},
  year={2022b},
  publisher={Wiley Online Library}
}

@article{heinze2024phases,
  title={Phases of methodological research in biostatistics---building the evidence base for new methods},
  author={Heinze, Georg and Boulesteix, Anne-Laure and Kammer, Michael and Morris, Tim P and White, Ian R},
  journal={Biometrical Journal},
  volume={66},
  number={1},
  pages={2200222},
  year={2024},
  publisher={Wiley Online Library}
}

@article{spiegelhalter1986monitoring,
  title={Monitoring clinical trials: conditional or predictive power?},
  author={Spiegelhalter, David J and Freedman, Laurence S and Blackburn, Patrick R},
  journal={Controlled clinical trials},
  volume={7},
  number={1},
  pages={8--17},
  year={1986},
  publisher={Elsevier}
}

@article{kunzmann2021review,
  title={A review of Bayesian perspectives on sample size derivation for confirmatory trials},
  author={Kunzmann, Kevin and Grayling, Michael J and Lee, Kim May and Robertson, David S and Rufibach, Kaspar and Wason, James MS},
  journal={The American Statistician},
  volume={75},
  number={4},
  pages={424--432},
  year={2021},
  publisher={Taylor \& Francis}
}

@misc{EMA_ExtControl_Event,
  author = {{European Medicines Agency}},
  title = {Workshop on the use of external controls for evidence generation in regulatory decision-making},
  howpublished = {\url{https://www.ema.europa.eu/en/events/workshop-use-external-controls-evidence-generation-regulatory-decision-making}},
  year = {2025},
  note = {Accessed: December 3, 2025}
}

@misc{EMA_2024_mwp,
  author = {{European Medicines Agency}},
  title = {Methodology Working Party Interested Parties Meeting},
  howpublished = {\url{https://www.ema.europa.eu/en/events/methodology-working-party-interested-parties-meeting}},
  year = {2024},
  note = {Accessed: January 27, 2026}
}

@article{berry2025statistical,
  title={Statistical innovations in clinical trial design with a focus on drug combinations, factorials, and other multiple therapy issues},
  author={Berry, Donald A},
  journal={The Journal of Prevention of Alzheimer's Disease},
  pages={100392},
  year={2025},
  publisher={Elsevier}
}

@misc{BBS_Past_Events,
  author = {{Basel Biometrics Society}},
  title = {Past events: agendas, slidedecks, recordings},
  howpublished = {\url{https://baselbiometrics.github.io/home/docs/events_past.html}},
  year = {2025},
  note = {Accessed: January 3, 2026}
}

@misc{EFSPI_RSW_Event,
  author = {{EFSPI}},
  title = {{EFSPI Regulatory Statistics Workshop}},
  howpublished = {\url{https://efspieurope.github.io/workshop/}},
  year = {2025},
  note = {Accessed: January 3, 2026}
}

@misc{ASA_RSW_Event,
  author = {{American Statistical Association}},
  title = {{ASA Biopharmaceutical Section Regulatory-Industry Statistics Workshop}},
  howpublished = {\url{https://www.amstat.org/meetings/asa-biopharmaceutical-section-regulatory-industry-statistics-workshop}},
  year = {2025},
  note = {Accessed: January 3, 2026}
}

@misc{ASAbiop_2025_wg,
  author = {{American Statistical Association}},
  title = {{ASA Biopharmaceutical Section Working Groups}},
  howpublished = {\url{https://community.amstat.org/biop/workinggroups}},
  year = {2025},
  note = {Accessed: January 3, 2026}
}

@misc{ISCB25_Event,
  author = {{ISCB}},
  title = {Statistical Thinking Across Borders:  Building Bridges and Expanding Horizons in Clinical Biostatistics},
  howpublished = {\url{https://iscb2025.info/}},
  year = {2025},
  note = {Accessed: January 3, 2026}
}

@misc{PSI_esig,
  author = {{PSI}},
  title = {{PSI / EFSPI - Special Interest Groups}},
  howpublished = {\url{https://psiweb.org/sigs-special-interest-groups/sigs}},
  year = {2025},
  note = {Accessed: January 3, 2026}
}

@misc{EIWG_esig,
  author = {{EFPIA and EFSPI}},
  title = {{Estimands Implementation Working Group (EIWG)}},
  howpublished = {\url{https://eiwg.github.io/eiwg\_webpage/}},
  year = {2026},
  note = {Accessed: January 20, 2026}
}

@misc{statmethwg_esig,
  author = {{EFSPI Statistical Methods Leaders}},
  title = {{EFSPI Statistical Methods Leaders}},
  howpublished = {\url{https://efspieurope.github.io/efspi/methods/methods_intro.html}},
  year = {2023},
  note = {Accessed: January 20, 2026}
}

@article{ferreira2025ai,
  title={{AI-driven drug discovery: A comprehensive review}},
  author={Ferreira, F{\'a}bio JN and Carneiro, Agnaldo S},
  journal={ACS omega},
  year={2025},
  publisher={ACS Publications}
}

@misc{fda2025roadmap,
  title={Roadmap to Reducing Animal Testing in Preclinical Safety Studies},
  author={{US FDA}},
  howpublished = {\url{https://www.fda.gov/files/newsroom/published/roadmap_to_reducing_animal_testing_in_preclinical_safety_studies.pdf}},
  year = {2025},
  note = {Accessed: January 3, 2026}
}

@misc{fda2025sig,
  title={{Project SignifiCanT: Statistics in cancer trials}},
  author={{US FDA}},
  howpublished = {\url{https://www.fda.gov/about-fda/oncology-center-excellence/project-significant-statistics-cancer-trials}},
  year = {2025},
  month = {10},
  note = {Accessed: January 3, 2026}
}

@misc{fda2025cidmp,
  title={Complex innovative trial design meeting program},
  author={{US FDA}},
  howpublished = {\url{https://www.fda.gov/drugs/development-resources/complex-innovative-trial-design-meeting-program}},
  year = {2025},
  note = {Accessed: January 3, 2026}
}

@article{liu2021evaluating,
  title={{Evaluating eligibility criteria of oncology trials using real-world data and AI}},
  author={Liu, Ruishan and Rizzo, Shemra and Whipple, Samuel and Pal, Navdeep and Pineda, Arturo Lopez and Lu, Michael and Arnieri, Brandon and Lu, Ying and Capra, William and Copping, Ryan and others},
  journal={Nature},
  volume={592},
  number={7855},
  pages={629--633},
  year={2021},
  publisher={Nature Publishing Group UK London}
}

@article{Best03042025,
author = {Nicky Best and Maxine Ajimi and Beat Neuenschwander and Gaëlle Saint-Hilary and Simon Wandel},
title = {{Beyond the classical Type I error: Bayesian metrics for Bayesian designs using informative priors}},
journal = {Statistics in Biopharmaceutical Research},
volume = {17},
number = {2},
pages = {183--196},
year = {2025},
publisher = {Taylor \& Francis}
}

@book{he2014practical,
  title={Practical considerations for adaptive trial design and implementation},
  author={He, Weili and Pinheiro, Jos{\'e} and Kuznetsova, Olga M},
  year={2014},
  publisher={Springer}
}

@article{russek2025supplementing,
  title={Supplementing Single-Arm Trials with External Control Arms—Evaluation of German Real-World Data},
  author={Russek, Martin and Peltner, Jonas and Haenisch, Britta},
  journal={Clinical Pharmacology \& Therapeutics},
  year={2025},
  publisher={Wiley Online Library}
}

@article{efspiwg1999qs,
  title={{Qualified statisticians in the european pharmaceutical industry:  Report of a European Federation of Statisticians in the Pharmaceutical Industry (EFSPI) working group}},
  author={{EFSPI Working Group}},
  journal= {Drug Information Journal},
  year={1999},
  volume={33},
  pages={407-415},
  publisher={Drug Information Association Inc.}
}

@misc{asabiop2025stat,
    author = {Fu, Haoda and Xia, H. Amy},
    title = {The Evolving Role of Statisticians in the Pharmaceutical Industry: Leveraging Advanced Statistical Analytics and Artificial Intelligence},
  howpublished = {\url{https://asabiopreport.substack.com/p/the-evolving-role-of-statisticians-ae6}},
  year = {2025},
  note = {Accessed: January 11, 2026}
}

@misc{ICHG2025E6,
  author = {{ICH}},
  title = {{ICH E6 (R3) Guideline for good clinical practice (GCP)}},
  howpublished = {\url{https://www.ema.europa.eu/en/documents/scientific-guideline/ich-e6-r3-guideline-good-clinical-practice-gcp-step-5_en.pdf}},
  month = {7},
  year = {2025},
  note = {"Step 5, adopted"},
  organization = {{International Council for Harmonisation of Technical Requirements for Pharmaceuticals for Human Use}}
}

@misc{ICHG2025E9,
  author = {{ICH}},
  title = {{ICH Topic E9 Statistical Principles for Clinical Trials}},
  howpublished = {\url{https://www.ema.europa.eu/en/documents/scientific-guideline/ich-e-9-statistical-principles-clinical-trials-step-5_en.pdf}},
  month = {9},
  year = {1997},
  note = {"Step 5, adopted"},
  organization = {{International Council for Harmonisation of Technical Requirements for Pharmaceuticals for Human Use}}
}

@misc{ICHG2019E9r1,
  author = {{ICH}},
  title = {{ICH Addendum on estimands and sensitivity analysis in clinical trials to the guideline on statistical practices for clinical trials, final version}},
  howpublished = {\url{https://database.ich.org/sites/default/files/E9-R1_Step4_Guideline_2019_1203.pdf}},
  month = {11},
  year = {2019},
  note = {"adopted"},
  organization = {{International Council for Harmonisation of Technical Requirements for Pharmaceuticals for Human Use}}
}

@misc{fdabayesian2026,
  author      = {{US FDA}},
  title       = {{Use of Bayesian Methodology in Clinical Trials of Drug and Biological Products: Draft Guidance for Industry}},
  howpublished = {\url{https://www.fda.gov/regulatory-information/search-fda-guidance-documents/use-bayesian-methodology-clinical-trials-drug-and-biological-products}},
  institution = {U.S. Department of Health and Human Services},
  year        = {2026},
  month       = {1}
}

@misc{fdaaireg2025,
  author      = {{US FDA}},
  title       = {{Considerations for the Use of Artificial Intelligence to Support Regulatory Decision-Making for Drug and Biological Products:   Draft Guidance for Industry and Other Interested Parties}},
  howpublished = {\url{https://www.fda.gov/regulatory-information/search-fda-guidance-documents/considerations-use-artificial-intelligence-support-regulatory-decision-making-drug-and-biological}},
  institution = {U.S. Department of Health and Human Services},
  year        = {2025},
  month       = {1}
}

@misc{fdarwd2023,
  author      = {{US FDA}},
  title       = {{Considerations for the Use of Real-World Data and Real-World Evidence To Support Regulatory Decision-Making for Drug and Biological Products:  Guidance for Industry}},
  howpublished = {\url{https://www.fda.gov/regulatory-information/search-fda-guidance-documents/considerations-use-real-world-data-and-real-world-evidence-support-regulatory-decision-making-drug}},
  institution = {U.S. Department of Health and Human Services},
  year        = {2023},
  month       = {8}
}

@article{behr2025opportunities,
  title={{Opportunities and challenges for AI-based analysis of RWD in pharmaceutical R\&D: a practical perspective}},
  author={Behr, Merle and Burghaus, Rolf and Diedrich, Christian and Lippert, J{\"o}rg},
  journal={KI-K{\"u}nstliche Intelligenz},
  volume={39},
  number={1},
  pages={7--18},
  year={2025},
  publisher={Springer}
}

@article{garczarek2023bayesian,
  title={Bayesian strategies in rare diseases},
  author={Garczarek, Ursula and Muehlemann, Natalia and Richard, Frank and Yajnik, Pranav and Russek-Cohen, Estelle},
  journal={Therapeutic innovation \& regulatory science},
  volume={57},
  number={3},
  pages={445--452},
  year={2023},
  publisher={Springer}
}

@article{bornkamp2024predicting,
  title={Predicting subgroup treatment effects for a new study: Motivations, results and learnings from running a data challenge in a pharmaceutical corporation},
  author={Bornkamp, Bj{\"o}rn and Zaoli, Silvia and Azzarito, Michela and Martin, Ruvie and M{\"u}ller, Carsten Philipp and Moloney, Conor and Capestro, Giulia and Ohlssen, David and Baillie, Mark},
  journal={Pharmaceutical statistics},
  volume={23},
  number={4},
  pages={495--510},
  year={2024},
  publisher={Wiley Online Library}
}

@article{bove2026sse,
  title={The statistical software revolution in pharmaceutical development: challenges and opportunities in open source},
  author={Saban{\'e}s Bov{\'e}, Daniel  and Seibold, Heidi and Boulesteix, Anne-Laure and Manitz, Juliane and Gasparini, Alessandro and G{\"u}nhan, Burak K and Boix, Oliver and Sch{\"u}ler, Armin and Fillinger, Sven and Nahnsen, Sven and others},
  journal={Drug Discovery Today},
  pages={104613},
  year={2026},
  publisher={Elsevier}
}

@article{rufibach2024survival,
  title={Survival analysis for AdVerse events with VarYing follow-up times (SAVVY): summary of findings and assessment of existing guidelines},
  author={Rufibach, Kaspar and Beyersmann, Jan and Friede, Tim and Schmoor, Claudia and Stegherr, Regina},
  journal={Trials},
  volume={25},
  number={1},
  pages={353},
  year={2024},
  publisher={Springer}
}

@article{asikanius2025clinical,
  title={Clinical trials with interim analyses: standardizing terminology to increase clarity},
  author={Asikanius, Elina and Hofner, Benjamin and Hampson, Lisa V and Wassmer, Gernot and Jennison, Christopher and Mielke, Tobias and Kunz, Cornelia Ursula and Rufibach, Kaspar},
  journal={Trials},
  volume={26},
  number={1},
  pages={247},
  year={2025},
  publisher={Springer}
}

@article{Vuong_2026estimands,
  title={Estimands and Their Implications for Evidence Synthesis for Oncology: A Simulation Study of Treatment Switching in Meta-Analysis},
  author={Vuong, Quang and Metcalfe, Rebecca K and Remiro-Az{\'o}car, Antonio and Gorst-Rasmussen, Anders and Keene, Oliver and Park, Jay JH},
  journal={Research Synthesis Methods},
  year={2026},
  volume={17}, 
DOI={10.1017/rsm.2025.10039}, number={1} 
}

@article{lanius2025realizing,
  title={Realizing the benefits of the estimand framework when reporting and communicating clinical trial results—some recommendations},
  author={Lanius, Vivian and Glocker, Barbara and L{\"o}sch, Christian and Bratton, Daniel J and Callegari, Francesca and Wright, Melanie and Rajam{\"a}ki, Suvi},
  journal={Trials},
  volume={26},
  number={1},
  pages={241},
  year={2025},
  publisher={Springer}
}

@article{dang2023causal,
  title={A causal roadmap for generating high-quality real-world evidence},
  author={Dang, Lauren E and Gruber, Susan and Lee, Hana and Dahabreh, Issa J and Stuart, Elizabeth A and Williamson, Brian D and Wyss, Richard and D{\'\i}az, Iv{\'a}n and Ghosh, Debashis and K{\i}c{\i}man, Emre and others},
  journal={Journal of Clinical and Translational Science},
  volume={7},
  number={1},
  pages={e212},
  year={2023},
  publisher={Cambridge University Press}
}

@article{ho2023current,
  title={The current landscape in biostatistics of real-world data and evidence: causal inference frameworks for study design and analysis},
  author={Ho, Martin and van der Laan, Mark and Lee, Hana and Chen, Jie and Lee, Kwan and Fang, Yixin and He, Weili and Irony, Telba and Jiang, Qi and Lin, Xiwu and others},
  journal={Statistics in Biopharmaceutical Research},
  volume={15},
  number={1},
  pages={43--56},
  year={2023},
  publisher={Taylor \& Francis}
}

@article{cortes2021strategic,
  title={Strategic leadership of innovation: a framework for future research},
  author={Cortes, Andres Felipe and Herrmann, Pol},
  journal={International Journal of Management Reviews},
  volume={23},
  number={2},
  pages={224--243},
  year={2021},
  publisher={Wiley Online Library}
}

@article{kulmann2016bic,
  title={The biostatistics innovation center at Bayer},
  author={Kulmann, Hermann and Muysers, Christoph and Dmitrienko, Alex and Roehmel, Joachim},
  journal={Biopharmaceutical Report},
  volume={23},
  number={3},
  year={2016},
  note={\url{https://higherlogicdownload.s3.amazonaws.com/AMSTAT/fa4dd52c-8429-41d0-abdf-0011047bfa19/UploadedImages/BIOP Report/2016_Fall_BR.pdf}}
}

@article{stallard2020,
  title={{Efficient adaptive designs for clinical trials of interventions for COVID-19}},
  author={Stallard, Nigel and Hampson, Lisa and Benda, Norbert and Brannath, Werner and Burnett, Thomas and Friede, Tim and Kimani, Peter K. and Koenig, Franz and Krisam, Johannes and Mozgunov, Pavel and Posch, Martin and Wason, James and Wassmer, Gernot and Whitehead, John and Williamson, S Faye and Zohar, Sara and Jaki, Thomas},
  journal={Statistics in Biopharmaceutical Research},
  volume={12},
  number={4},
  year={2020},
  url={https://pmc.ncbi.nlm.nih.gov/articles/PMC8011600/},
}

@article{roychoudhury2023robust,
  title={Robust design and analysis of clinical trials with nonproportional hazards: a straw man guidance from a cross-pharma working group},
  author={Roychoudhury, Satrajit and Anderson, Keaven M and Ye, Jiabu and Mukhopadhyay, Pralay},
  journal={Statistics in Biopharmaceutical Research},
  volume={15},
  number={2},
  pages={280--294},
  year={2023},
  publisher={Taylor \& Francis}
}

@book{christensen2015innovator,
  title = {The innovator's dilemma: when new technologies cause great firms to fail},
  author = {Christensen, Clayton M.},
  year = {2015},
  publisher = {Harvard Business Review Press}
}

@article{dallow2018,
  title={Better decision making in drug development through adoption of formal prior elicitation},
  author={Dallow, N. and Best, N. and Montague, T.H.},
  journal={Pharmaceutical Statistics},
  volume={17},
  number={4},
  pages={301--316},
  year={2018},
  publisher={Wiley Online Library}
}

@article{bornkamp2009mcpmod,
  title={{MCPMod: An R package} for the design and analysis of dose-finding studies},
  author={Bornkamp, Bj{\"o}rn and Pinheiro, Jos{\'e} and Bretz, Frank},
  journal={Journal of Statistical Software},
  volume={29},
  pages={1--23},
  year={2009}
}

@article{neuenschwander08,
  title={Critical aspects of the Bayesian approach to phase I cancer trials},
  author={Neuenschwander, Beat and Branson, Michael and Gsponer, Thomas},
  journal={Statistics in Medicine},
  volume={27},
  number={13},
  pages={2420--2439},
  year={2008}
}

\end{document}